\documentclass[preprint]{JHEP3} 

\JHEPspecialurl{http://jhep.sissa.it/JOURNAL/JHEP3.tar.gz}
\usepackage{epsfig,multicol}
\usepackage{axodraw}

\newcommand\fverb{\setbox\fverbbox=\hbox\bgroup\verb}

\title{Primary Feynman rules 
to calculate the  $\epsilon-$dimensional integrand
of any 1-loop amplitude}

\author{R. Pittau\\
       Departamento de F\'{i}sica Te\'orica y del Cosmos y CAFPE
       Universidad de Granada, E-18071 Granada, Spain.\\
       E-mail: \email{pittau@ugr.es}}

\preprint{} 

\abstract{
When using  dimensional regularization/reduction the
$\epsilon$-dimensional numerator of the 1-loop Feynman diagrams
gives rise to rational contributions.
I list the set of fundamental rules that allow the extraction 
of such terms at the integrand level in any theory
containing scalars, vectors and fermions, such as the electroweak 
standard model, QCD and SUSY.   
}
\keywords{NLO, radiative corrections, electroweak model, QCD}

\begin{document}

\newcounter{im}
\setcounter{im}{0}
\newcommand{\exampleSp}{\stepcounter{im}\includegraphics[scale=0.9]{SpinorExamples_\arabic{im}.eps}}
\newcommand{\myindex}[1]{\label{com:#1}\index{{\tt #1} & pageref{com:#1}}}
\renewcommand{\topfraction}{1.0}
\renewcommand{\bottomfraction}{1.0}
\renewcommand{\textfraction}{0.0}
\newcommand{\nn}{\nonumber \\}
\newcommand{\eqn}[1]{eq.~\ref{eq:#1}}
\newcommand{\be}{\begin{equation}}
\newcommand{\ee}{\end{equation}}
\newcommand{\ba}{\begin{array}}
\newcommand{\ea}{\end{array}}
\newcommand{\bea}{\begin{eqnarray}}
\newcommand{\eea}{\end{eqnarray}}
\newcommand{\bqa}{\begin{eqnarray}}
\newcommand{\eqa}{\end{eqnarray}}
\newcommand{\nl}{\nonumber \\}
\def\db#1{\bar D_{#1}}
\def\zb#1{\bar Z_{#1}}
\def\d#1{D_{#1}}
\def\tld#1{\tilde {#1}}
\def\slh#1{\rlap / {#1}}
\def\eqn#1{eq.~(\ref{#1})}
\def\eqns#1#2{fqs.~(\ref{#1}) and~(\ref{#2})}
\def\eqnss#1#2{fqs.~(\ref{#1})-(\ref{#2})}
\def\fig#1{fig.~{\ref{#1}}}
\def\figs#1#2{figs.~\ref{#1} to~\ref{#2}}
\def\figss#1#2{figs.~\ref{#1} and~\ref{#2}}
\def\sec#1{fection~{\ref{#1}}}
\def\app#1{appendix~\ref{#1}}
\def\tab#1{table~\ref{#1}}
\def\cg{c_\Gamma}
\newcommand{\bfig}{\begin{center}\begin{picture}}
\newcommand{\efig}[1]{\end{picture}\\{\small #1}\end{center}}
\newcommand{\flin}[2]{\ArrowLine(#1)(#2)}
\newcommand{\ghlin}[2]{\DashArrowLine(#1)(#2){5}}
\newcommand{\wlin}[2]{\DashLine(#1)(#2){2.5}}
\newcommand{\zlin}[2]{\DashLine(#1)(#2){5}}
\newcommand{\glin}[3]{\Photon(#1)(#2){2}{#3}}
\newcommand{\gluon}[3]{\Gluon(#1)(#2){5}{#3}}
\newcommand{\lin}[2]{\Line(#1)(#2)}
\newcommand{\sof}{\SetOffset}


\section{\label{sec:1}Introduction}
New techniques~\cite{Bern:1994zx,Bern:1994cg,Britto:2004nc,Ossola:2006us,Forde:2007mi,Ossola:2007ax,Berger:2008sj,Giele:2008bc,Ellis:2009zw,vanHameren:2009dr,Berger:2009ep,Hirschi:2011pa,Bevilacqua:2011xh,Hirschi:2011rb} for computing 1-loop corrections 
led to a NLO revolution~\cite{Salam:2011bj,Ellis:2011cr}, that, directly or indirectly, has allowed an impressing improvement in our ability to predict physical observables at the NLO accuracy~\cite{Binoth:2008kt,Ellis:2008qc,Berger:2009zg,Bredenstein:2009aj,Bevilacqua:2009zn,Bredenstein:2010rs,Bevilacqua:2010ve,Binoth:2010ra,Melnikov:2010iup,Berger:2010zx,Denner:2010jp,Denner:2010jp,Bevilacqua:2010qb,Melia:2011dw,Frederix:2011zi,Campanario:2011ud,Frederix:2011qg,Arnold:2011wj,Ita:2011wn,Bevilacqua:2011hy,Frederix:2011ss,Frederix:2011ig}.
Basically all new methods need a special treatment of the contributions that
are not proportional to the scalar 1-loop functions, 
the so called rational terms~\cite{Binoth:2006hk,Ossola:2008xq,Bredenstein:2008zb,Campanario:2011cs}. That is achieved, in  Unitarity and Generalized Unitarity methods, 
by computing the entire amplitude in different numbers of 
space-time dimensions~\cite{Giele:2008ve}, or via bootstrapping 
techniques~\cite{Bern:2005cq,Badger:2007si}, or through
$d$-dimensional cuts~\cite{Badger:2008cm}. The  Ossola-Papadopoulos-Pittau (OPP) approach of reference~\cite{Ossola:2006us} requires, instead, the computation (once for all for the theory at hand) of a special set of tree level Feynman rules~\cite{Draggiotis:2009yb,Garzelli:2009is,Garzelli:2010qm,Garzelli:2010fq,Shao:2011tg} up to 4-point interactions \footnote{The contributions to higher-point functions vanish because of UV finiteness.}.

When using dimensional regularization/reduction, the origin of
such terms lies in the $\epsilon$-dimensional numerator of the 1-loop 
Feynman diagrams \footnote{Another contribution, which is however 
directly linkable to the cut-constructible part of the amplitude
~\cite{Ossola:2007bb}, is generated by the  $\epsilon$-dimensional 1-loop denominators.}. To be more specific, let us consider the general 
expression for the integrand of a generic $m$-point
one-loop (sub-) amplitude
\bqa
\label{eq:1}
\bar A(\bar q)= \frac{\bar N(\bar q)}{\db{0}\db{1}\cdots \db{m-1}}\,,~~~
\db{i} = ({\bar q} + p_i)^2-m_i^2\,,
\eqa
where ${\bar q}$ is the integration momentum and
\bqa
\bar q^2 &=& q^2 +  \tld{q}^2 ~\equiv~q^2 -  \mu^2 \,. 
\eqa
In the previous expressions and in all the following ones, a bar denotes objects living
in $d=~4+\epsilon$ dimensions, whereas a tilde represents $\epsilon$-dimensional
quantities.

The numerator function $\bar{N}(\bar q)$ can be 
split into a $4$-dimensional plus an $\epsilon$-dimensional part
\bqa
\label{eq:split}
\bar{N}(\bar q) = N(q) + \tld{N}(\mu^2,q,\epsilon)\,.
\eqa
$N(q)$ lives in $4$ dimensions, while $\tld{N}(\mu^2,q,\epsilon)$, which 
originates from the splitting of $d$-dimensional objects
\bqa
\label{qandg}
\bar q                 &=& q + \tld{q}\,, \nl
\bar \gamma_{\bar \mu} &=&  \gamma_{\mu}+ \tld{\gamma}_{\tld{\mu}}\,,\nl
 \bar g^{\bar \mu \bar \nu}  &=&  g^{\mu \nu}+  \tld{g}^{\tld{\mu} \tld{\nu}}\,,
\eqa
gives rise to a rational piece called ${\rm R_2}$ in the OPP language:
\bqa
\label{eqr2}
{\rm R_2}  \Bigl |_{\rm HV} = \frac{1}{(2 \pi)^4}\int d^d\,\bar q
\frac{\tld{N}(\mu^2,q,\epsilon)}{\db{0}\db{1}\cdots \db{m-1}} \,,
\eqa 
with
\bqa
\int d^d\,\bar q~=~\int d^4q \int d^\epsilon \mu\,.
\eqa
${\rm R_2}$ has a pure ultraviolet origin~\cite{Binoth:2006hk,Bredenstein:2008zb}, so that~\eqn{eqr2} also holds when infrared/collinear divergences are present in the loop integrals.  
It can be shown~\cite{Ossola:2008xq,Giele:2008ve} that
$\tld{N}(\mu^2,q,\epsilon)$ is polynomial in $\mu^2$ 
and at most linear in $\epsilon$, and that the $\epsilon$ dependence can be
reabsorbed in the regularization scheme~\cite{Signer:2008va}. 
Therefore, beside
\eqn{eqr2}, which defines ${\rm R_2}$ in the 't Hooft-Veltman (HV) scheme,
a Four Dimensional Helicity scheme (FDH) can be used in which 
the $\epsilon$ dependence in the numerator function is discarded 
before integration
\bqa
\label{eqr2fdh}
{\rm R_2} \Bigl |_{\rm FDH} =  \frac{1}{(2 \pi)^4}\int d^d\,\bar q
\frac{\tld{N}(\mu^2,q,\epsilon= 0)}{\db{0}\db{1}\cdots \db{m-1}} \,.
\eqa
As for the virtual part of the NLO corrections, FDH is equivalent to Dimensional Reduction~\cite{Stockinger:2005gx}.

It is clear that explicitly using the rules in~\eqn{qandg} allows an analytic extraction, Feynman diagram by Feynman diagram, of the coefficients of the various powers of $\mu^2$  and $\epsilon$.
For example, the {\tt GoSam}~\cite{Mastrolia:2010nb,Cullen:2011aw} approach achieves this {\em on the fly}, by linking to algebraical manipulation programs  providing the necessary algebra when building the amplitude.
The case of QCD is particularly simple, and computations based on a standard Passarino-Veltman~\cite{Passarino:1978jh} decomposition are relatively easy~\cite{Xiao:2006vr,Badger:2011yu}. In addition, for gluonic amplitudes, a super-symmetric decomposition relates the contribution of the rational
terms to a scalar massive gluon running in the loop~\cite{Badger:2010nx}, which can also be computed by using massive Cachazo-Svr\v{c}ek-Witten Feynman rules~\cite{Boels:2007pj,NigelGlover:2008ur,Elvang:2011ub}.

All the approaches mentioned so far have advantages and drawbacks. It is however beyond doubt that it would be desirable to have a way to compute ${\rm R_2}$ {\em independent} of the theory at hand, {\em four dimensional} and
{\em not requiring the use of analytical manipulations}.
In this paper, I provide such a method in the form of primary Feynman rules, which reproduce the polynomial dependence on $\mu^2$ of the integrand in~\eqn{eqr2fdh} \footnote{The conversion to the HV 
scheme of~\eqn{eqr2} is presented in appendix~\ref{sec:app}.} when working
in the renormalizable gauge. As  will be explained in the next section, such rules are 
uniquely determined by reading the original propagators and vertices
of the theory, and solely depend on their Lorentz structure.
They can therefore be used as any other Feynman rules 
by programs such as {\tt MadLoop}~\cite{Hirschi:2011pa}, 
{\tt HELAC-NLO}~\cite{Bevilacqua:2011xh}, 
{\tt FeynRules}~\cite{Christensen:2008py} or within a methods such as
{\tt Open Loops}~\cite{Cascioli:2011va} to automatically 
generate ${\rm R_2}$. 

 This paper is organized as follows: in section~\ref{sec:2} 
I give an introductory discussion and fix my conventions.
Section~\ref{sec:3}  introduces vectors, scalars and their 
mutual interactions. 
 The fermionic case is more delicate, and it is discussed in 
section~\ref{sec:4}. In section~\ref{sec:5} the generic formulae
are specialized to the QCD case and, in section~\ref{sec:6}, 
I draw my conclusions.
Further details are reported in appendix~\ref{sec:app}.

\section{\label{sec:2}Preliminaries and conventions}
I am interested in reconstructing the polynomial dependence on
$\mu^2$ of the $\epsilon$-dimensional numerator in~\eqn{eqr2fdh}.
In a renormalizable gauge the relevant expansion is
\bqa
\label{eq:expn}
\tilde N(\mu^2,q,\epsilon= 0)= \sum_{j=1}^2 (\mu^2)^j \,c_j(q)\,.
\eqa
As already stated in section~\ref{sec:1}, an explicit dependence on
$\epsilon$, such as that implied by~\eqn{eqr2}, can always be reproduced 
by a change of regularization scheme. The translation
rules relevant for QCD and for the electroweak standard 
model are given in appendix~\ref{sec:app}.

By inserting~\eqn{eq:expn} into~\eqn{eqr2fdh} the relevant integrals 
are of the kind 
\bqa
\frac{1}{(2 \pi)^4}\int d^d\,\bar q
\frac{(\mu^2)^j q_{\mu_1} \cdots q_{\mu_r}}{\db{0}\db{1}\cdots \db{m-1}}
~~~~{\rm with}~~~~m \leq 4\,,~0 < j\le 2\,,~r\le 2\,,
\eqa
which give a non-vanishing contribution, in the limit $\epsilon \to 0$, only 
when~$4+2j+r-2m \geq 0$. They are computed in ~\cite{Ossola:2007bb}.

Powers of $\mu^2$ in ~\eqn{eq:expn} are generated by the contraction 
of the integration momentum with itself only in the presence of 
vectors and fermions, which are indeed the 
only particles bringing the integration momentum in the numerator. 
For each vector field $V_\alpha$, I therefore introduce a scalar
field $\hat V$, whose propagator corresponds to the propagation 
of its $\epsilon$-dimensional components. 
In the same way, for any fermion $F$, I introduce 
a fermionic field $\hat F$ and its corresponding propagator. 
In the following, $\hat V$ and $\hat F$ are called $\epsilon$-particles
and their propagators are graphically represented by a dashed line 
and a dashed arrow,  respectively.

$\epsilon$-particles interact among themselves and with the 
particles of the original theory, bringing an explicit
dependence on $\mu$ and are only allowed to circulate in the loop.
I call $\epsilon$-vertices the 
special vertices involving $\epsilon$-particles, and $\epsilon$-diagrams the
Feynman diagrams that contain $\epsilon$-vertices.
Given the fact that the only possible mechanisms for generating powers of 
$\mu^2$ are those illustrated in~\eqn{qandg}, it is easy to determine the 
$\epsilon$-vertices by simply looking at the Feynman rules of the original 
theory. In particular, the Lorentz and color structures are completely dictated by the nature of the particles entering the $\epsilon$-vertices, so that, in general, only the relative phases of $\mu$ and/or $\sqrt{\mu}$ have to be determined. I fixed them by explicitly writing down all possible classes of 2,3, and 4-point Feynman diagrams involving scalars, vectors and fermions and by requiring that the $\epsilon$-vertices reproduce the results obtained with an explicit calculation of ${\rm R_2}$. It should be possible to determine the $\epsilon$-vertices directly from the original Lagrangian, by splitting it into 4 and $\epsilon$ dimensional parts, but I did not choose such an approach. 
In the following two sections, I list all possibilities involving
vectors, scalars and fermions. I do it in a completely generic way, in the sense that any particular model can be implemented just by giving a specific value 
to the constants in front of the listed Lorentz structures.
For example, the electroweak model is obtained by fixing them 
according to reference~\cite{Denner:1991kt} and the minimal supersymmetric standard model by using the rules in~\cite{Rosiek:1995kg}.

\section{\label{sec:3}Vectors, scalars and their interactions}
The propagator of a scalar $\epsilon$-particle associated with a
vector field is given in~\fig{fig:1}. 
From the original three-vector vertex, two corresponding
$\epsilon$-vertices are derived, as illustrated in~\fig{fig:2}. 
To determine 
the sign of the first $\epsilon$-vertex, one should keep track
of the flow of the loop momentum $q$.
The original four-vector vertex gives rise to the two $\epsilon$-vertices 
of~\fig{fig:3}, while two-vector-one-scalar, one-vector-two-scalar
and two-vector-two-scalar vertices generate just one 
$\epsilon$-vertex each, as shown in~\figs{fig:4}{fig:6}.
  
\begin{figure}[t]
\begin{center}

\fbox{
\begin{picture}(80,60)
\SetScale{0.8}
\SetOffset(55,22)
\Photon(-35,-5)(-5,-5){2}{5}
\Photon(-5,25)(-35,25){2}{5}
\Photon(-35,25)(-35,-5){2}{5}
\Photon(-5,-5)(-5,25){2}{5}
\Photon(-35,-5)(-55,-25){2}{5}
\Photon(-5,-5)(15,-25){2}{5}
\Photon(-35,25)(-55,45){2}{5}
\Photon(-5,25)(15,45){2}{5}
\end{picture}
}

\begin{picture}(300,280)
\SetScale{0.8}
\SetOffset(80,220)
\Photon(-35,-5)(-5,-5){2}{5}
\Photon(-5,25)(-35,25){2}{5}
\DashLine(-35,25)(-35,-5){2}
\DashLine(-5,-5)(-5,25){2}
\Photon(-35,-5)(-55,-25){2}{5}
\Photon(-5,-5)(15,-25){2}{5}
\Photon(-35,25)(-55,45){2}{5}
\Photon(-5,25)(15,45){2}{5}

\SetOffset(245,220)
\DashLine(-35,-5)(-5,-5){2}
\DashLine(-5,25)(-35,25){2}
\Photon(-35,25)(-35,-5){2}{5}
\Photon(-5,-5)(-5,25){2}{5}
\Photon(-35,-5)(-55,-25){2}{5}
\Photon(-5,-5)(15,-25){2}{5}
\Photon(-35,25)(-55,45){2}{5}
\Photon(-5,25)(15,45){2}{5}

\SetOffset(0,150)
\Photon(-35,-5)(-5,-5){2}{5}
\Photon(-5,25)(-35,25){2}{5}
\DashLine(-35,25)(-35,-5){2}
\Photon(-5,-5)(-5,25){2}{5}
\Photon(-35,-5)(-55,-25){2}{5}
\Photon(-5,-5)(15,-25){2}{5}
\Photon(-35,25)(-55,45){2}{5}
\Photon(-5,25)(15,45){2}{5}

\SetOffset(110,150)
\Photon(-35,-5)(-5,-5){2}{5}
\DashLine(-5,25)(-35,25){2}
\Photon(-35,25)(-35,-5){2}{5}
\Photon(-5,-5)(-5,25){2}{5}
\Photon(-35,-5)(-55,-25){2}{5}
\Photon(-5,-5)(15,-25){2}{5}
\Photon(-35,25)(-55,45){2}{5}
\Photon(-5,25)(15,45){2}{5}

\SetOffset(220,150)
\Photon(-35,-5)(-5,-5){2}{5}
\Photon(-5,25)(-35,25){2}{5}
\Photon(-35,25)(-35,-5){2}{5}
\DashLine(-5,-5)(-5,25){2}
\Photon(-35,-5)(-55,-25){2}{5}
\Photon(-5,-5)(15,-25){2}{5}
\Photon(-35,25)(-55,45){2}{5}
\Photon(-5,25)(15,45){2}{5}

\SetOffset(330,150)
\DashLine(-35,-5)(-5,-5){2}
\Photon(-5,25)(-35,25){2}{5}
\Photon(-35,25)(-35,-5){2}{5}
\Photon(-5,-5)(-5,25){2}{5}
\Photon(-35,-5)(-55,-25){2}{5}
\Photon(-5,-5)(15,-25){2}{5}
\Photon(-35,25)(-55,45){2}{5}
\Photon(-5,25)(15,45){2}{5}

\SetOffset(0,80)
\Photon(-35,-5)(-5,-5){2}{5}
\DashLine(-5,25)(-35,25){2}
\DashLine(-35,25)(-35,-5){2}
\Photon(-5,-5)(-5,25){2}{5}
\Photon(-35,-5)(-55,-25){2}{5}
\Photon(-5,-5)(15,-25){2}{5}
\Photon(-35,25)(-55,45){2}{5}
\Photon(-5,25)(15,45){2}{5}

\SetOffset(110,80)
\Photon(-35,-5)(-5,-5){2}{5}
\DashLine(-5,25)(-35,25){2}
\Photon(-35,25)(-35,-5){2}{5}
\DashLine(-5,-5)(-5,25){2}
\Photon(-35,-5)(-55,-25){2}{5}
\Photon(-5,-5)(15,-25){2}{5}
\Photon(-35,25)(-55,45){2}{5}
\Photon(-5,25)(15,45){2}{5}

\SetOffset(220,80)
\DashLine(-35,-5)(-5,-5){2}
\Photon(-5,25)(-35,25){2}{5}
\Photon(-35,25)(-35,-5){2}{5}
\DashLine(-5,-5)(-5,25){2}
\Photon(-35,-5)(-55,-25){2}{5}
\Photon(-5,-5)(15,-25){2}{5}
\Photon(-35,25)(-55,45){2}{5}
\Photon(-5,25)(15,45){2}{5}

\SetOffset(330,80)
\DashLine(-35,-5)(-5,-5){2}
\Photon(-5,25)(-35,25){2}{5}
\DashLine(-35,25)(-35,-5){2}
\Photon(-5,-5)(-5,25){2}{5}
\Photon(-35,-5)(-55,-25){2}{5}
\Photon(-5,-5)(15,-25){2}{5}
\Photon(-35,25)(-55,45){2}{5}
\Photon(-5,25)(15,45){2}{5}

\SetOffset(0,10)
\Photon(-35,-5)(-5,-5){2}{5}
\DashLine(-5,25)(-35,25){2}
\DashLine(-35,25)(-35,-5){2}
\DashLine(-5,-5)(-5,25){2}
\Photon(-35,-5)(-55,-25){2}{5}
\Photon(-5,-5)(15,-25){2}{5}
\Photon(-35,25)(-55,45){2}{5}
\Photon(-5,25)(15,45){2}{5}

\SetOffset(110,10)
\DashLine(-35,-5)(-5,-5){2}
\DashLine(-5,25)(-35,25){2}
\Photon(-35,25)(-35,-5){2}{5}
\DashLine(-5,-5)(-5,25){2}
\Photon(-35,-5)(-55,-25){2}{5}
\Photon(-5,-5)(15,-25){2}{5}
\Photon(-35,25)(-55,45){2}{5}
\Photon(-5,25)(15,45){2}{5}

\SetOffset(220,10)
\DashLine(-35,-5)(-5,-5){2}
\Photon(-5,25)(-35,25){2}{5}
\DashLine(-35,25)(-35,-5){2}
\DashLine(-5,-5)(-5,25){2}
\Photon(-35,-5)(-55,-25){2}{5}
\Photon(-5,-5)(15,-25){2}{5}
\Photon(-35,25)(-55,45){2}{5}
\Photon(-5,25)(15,45){2}{5}

\SetOffset(330,10)
\DashLine(-35,-5)(-5,-5){2}
\DashLine(-5,25)(-35,25){2}
\DashLine(-35,25)(-35,-5){2}
\Photon(-5,-5)(-5,25){2}{5}
\Photon(-35,-5)(-55,-25){2}{5}
\Photon(-5,-5)(15,-25){2}{5}
\Photon(-35,25)(-55,45){2}{5}
\Photon(-5,25)(15,45){2}{5}

\end{picture}

\vspace{2cm}

\fbox{
\begin{picture}(80,50)
\SetScale{0.7}

\SetOffset(25,22)
\Photon(35,5)(0,20){2}{5}
\Photon(0,-10)(35,5){2}{5}
\Photon(0,20)(0,-10){2}{5}
\Photon(70,35)(35,5){2}{5}
\Photon(35,5)(70,-25){2}{5}
\Photon(0,20)(-35,35){2}{5}
\Photon(-35,-25)(0,-10){2}{5}
\end{picture}
}

\begin{picture}(300,70)
\SetScale{0.7}
\SetOffset(50, 20)
\Photon(35,5)(0,20){2}{5}
\Photon(0,-10)(35,5){2}{5}
\DashLine(0,20)(0,-10){2}
\Photon(70,35)(35,5){2}{5}
\Photon(35,5)(70,-25){2}{5}
\Photon(0,20)(-35,35){2}{5}
\Photon(-35,-25)(0,-10){2}{5}

\SetOffset(220, 20)
\DashLine(35,5)(0,20){2}
\DashLine(0,-10)(35,5){2}
\Photon(0,20)(0,-10){2}{5}
\Photon(70,35)(35,5){2}{5}
\Photon(35,5)(70,-25){2}{5}
\Photon(0,20)(-35,35){2}{5}
\Photon(-35,-25)(0,-10){2}{5}

\end{picture}

\end{center}
\caption{
\label{fig:6bis} 
Examples of $\epsilon$-diagrams. In the two boxes I draw the 
original diagrams. The $\epsilon$-diagrams below each box 
reconstruct the complete $\mu^2$ dependence of the integrand of
the corresponding diagram, provided the same rooting for the loop momentum 
is chosen in each of them.}
\end{figure}
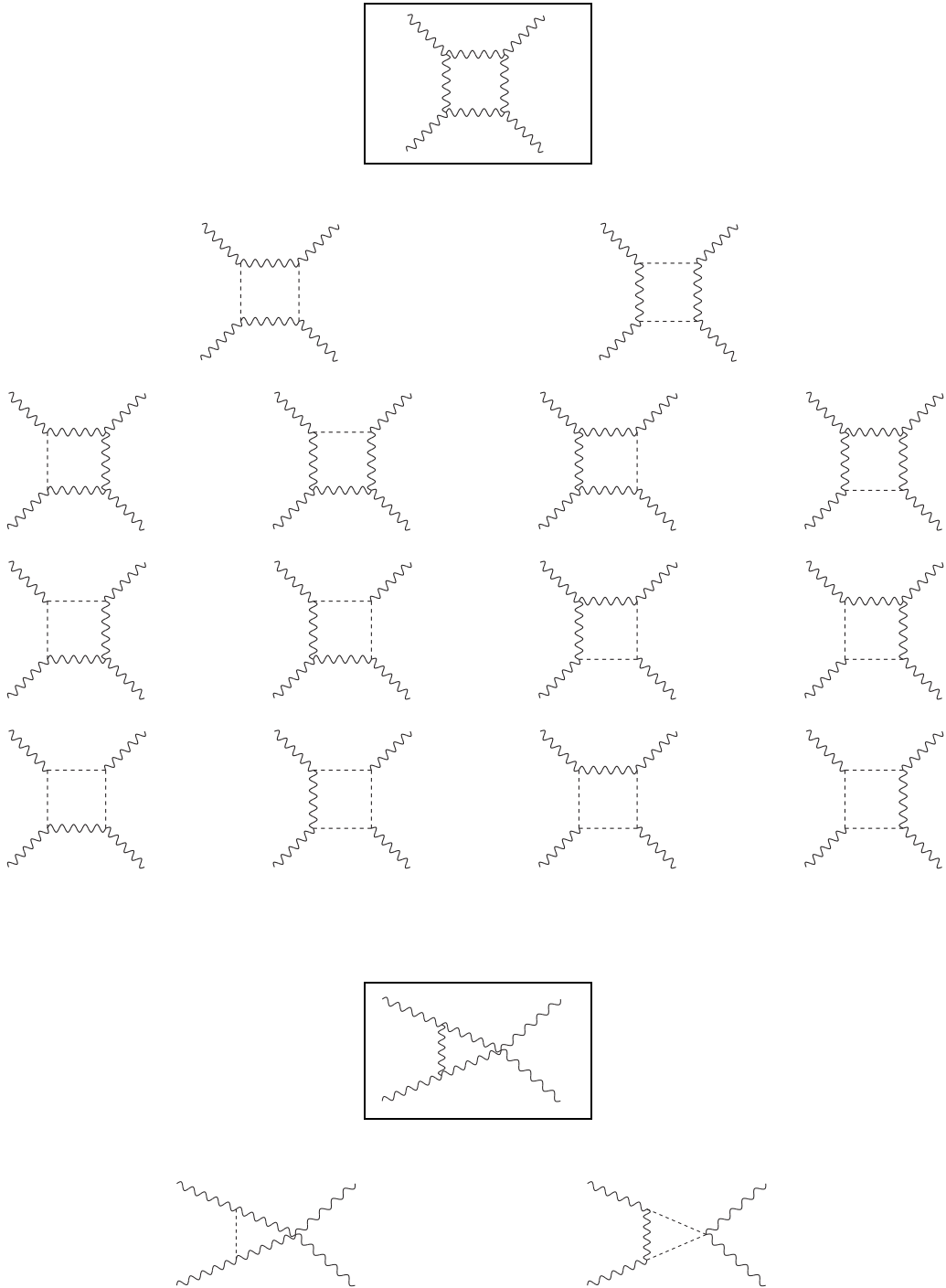

Although the $\epsilon$-vertices can be used in any stage of the calculation
to compute  $\epsilon$-diagrams reproducing the coefficients $c_j(q)$ of~\eqn{eq:expn}, it may be useful and illuminating to consider them in strict connection with the original Feynman diagrams.
With this kind of reasoning, one may associate to any 1-loop Feynman diagram contributing to the amplitude under study, a set of
$\epsilon$-diagrams which fully reconstruct its $\mu^2$ dependence. 
In particular, when using the same rooting for the loop momentum in each of them, this reconstruction even holds at the integrand level.
Furthermore, given the fact that the $\epsilon$-vertices bring known powers of $\mu$ into the $\epsilon$-diagrams, one can easily disentangle sub-sets 
generating specific powers of $\mu^2$ in~\eqn{eq:expn}. 
A concrete example is given in~\fig{fig:6bis} for two particular 
diagrams contributing to a $VV \to VV$ scattering, where 
the sum of the two $\epsilon $-diagrams in the second line yields 
the term $\mu^4\,c_2(q)$ corresponding to the original box diagram.

\begin{figure}[t]
\begin{center}
\fbox{
\begin{picture}(250,60)
\SetOffset(90,18)
\Text(-15,27)[l]{$ p $}
\LongArrow(-25,20)(0,20)
\Text(-65,10)[l]{$V_{\alpha} $}
\Photon(-50,10)(20,10){1}{9}
\Text(25,10)[l]{$V_{\beta} $}
\Text(55,8)[l]{$=~~\displaystyle -i \frac{g_{\alpha\beta}}{p^2-M^2_V} $}
\end{picture}
}
\begin{picture}(250,50)
\SetOffset(90,10)
\Text(-65,12)[l]{$ \hat{V} $}
\DashLine(-50,10)(20,10){2}
\Text(25,12)[l]{$ \hat{V} $}
\Text(55,10)[l]{$=~~\displaystyle -i \frac{1}{p^2-M^2_V} $}
\end{picture}
\caption{\label{fig:1} 
Propagator of a vector particle (box on the top) and propagator of its corresponding
scalar $\epsilon$-particle.}
\end{center}
\end{figure}

\begin{figure}[t]
\begin{center}
\fbox{
\begin{picture}(405,60)
\SetOffset(65,-174)\Photon(-35,200)(0,200){1}{7}
\Text(-40,208)[r]{$ V_{1\alpha} $}
\Text(-19,213)[l]{$ p_1 $}
\LongArrow(-20,205)(-10,205)
\Photon(0,200)(25,215){1}{7}
\Text(47,225)[tr]{$ V_{2\beta} $}
\Text(5,220)[l]{$ p_2 $}
\LongArrow(13,215)(5,210)
\Photon(25,190)(0,200){1}{7}
\Text(5,185)[l]{$ p_3 $}
\LongArrow(15,188)(5,192)
\Text(47,182)[br]{$  V_{3\gamma} $}
\Text(55,201)[l]{$=~~\displaystyle -ie\,C\,
\left[g_{\alpha\beta} (p_2-p_1)_\gamma
+g_{\beta\gamma} (p_3-p_2)_\alpha
+g_{\gamma\alpha}(p_1-p_3)_\beta
\right]
$}
\end{picture}
 }
\begin{picture}(405,140)
\SetOffset(65,-110)
\DashLine(-35,200)(0,200){2}
\Text(-45,208)[r]{$ \hat{V}_1 $}
\Photon(0,200)(25,215){1}{7}
\Text(47,225)[tr]{$ V_{2\beta}  $}
\Photon(25,190)(0,200){1}{7}
\Text(47,182)[br]{$  V_{3\gamma} $}
\Text(5,183)[l]{$ \pm q $}
\LongArrow(15,188)(5,192)
\Text(55,201)[l]{$=~~$}
\SetOffset(200,-110)
\DashLine(-35,200)(0,200){2}
\Text(-44,208)[r]{$ \hat{V}_1 $}
\Photon(0,200)(25,215){1}{7}
\Text(47,225)[tr]{$ V_{2\beta}  $}
\Photon(25,190)(0,200){1}{7}
\Text(47,182)[br]{$  V_{3\gamma} $}
\Text(-6,218)[l]{$ \pm q $}
\LongArrow(5,210)(13,215)
\Text(55,201)[l]{$=~~\displaystyle -ie\,C\,(\pm i \mu)\,g_{\beta\gamma} $}
\SetOffset(65,-180)
\Photon(-35,200)(0,200){1}{7}
\Text(-40,208)[r]{$ V_{1\alpha} $}
\DashLine(0,200)(25,215){2}
\Text(47,225)[tr]{$ \hat{V}_2 $}
\Text(5,220)[l]{$ p_2 $}
\LongArrow(13,215)(5,210)
\DashLine(25,190)(0,200){2}
\Text(5,185)[l]{$ p_3 $}
\LongArrow(15,188)(5,192)
\Text(47,182)[br]{$  \hat{V}_3 $}
\Text(55,201)[l]{$=~~\displaystyle -ie\,C\,(p_3-p_2)_{\alpha} $}
\end{picture}
\caption{\label{fig:2} 
Three-vector vertex (box on the top) and its corresponding $\epsilon$-vertices.
$q$ represents the flow of the loop momentum.
}
\end{center}
\end{figure}

\begin{figure}[t]
\begin{center}
\fbox{
\begin{picture}(350,90)
\SetOffset(80,-155)
\Text(-47,235)[]{$ V_{1\alpha} $}
\Photon(-35,230)(0,200){1}{7}
\Text(47,232)[]{$ V_{3\gamma} $}
\Photon(0,200)(35,230){1}{7}
\Text(-47,165)[]{$ V_{2\beta} $}
\Photon(-35,170)(0,200){1}{7}
\Text(47,165)[]{$ V_{4\delta} $}
\Photon(0,200)(35,170){1}{7}
\Text(90,200)[l]{$=~~\displaystyle ie^2C\,\left[2g_{\alpha\beta}g_{\gamma\delta}-g_{\beta\gamma}g_{\alpha\delta}-g_{\alpha\gamma}g_{\beta\delta}\right] $}

\end{picture}
}
\begin{picture}(350,220)
\SetOffset(80,-40)
\Text(-47,235)[]{${\hat V}_1 $}
\DashLine(-35,230)(0,200){2}
\Text(47,232)[]{$ V_{3\gamma} $}
\Photon(0,200)(35,230){1}{7}
\Text(-47,165)[]{$ \hat{V}_2 $}
\DashLine(-35,170)(0,200){2}
\Text(47,165)[]{$ V_{4\delta} $}
\Photon(0,200)(35,170){1}{7}
\Text(90,200)[l]{$=~~\displaystyle ie^2\,C\,(2g_{\gamma\delta}) $}
\SetOffset(80,-150)
\Text(-47,235)[]{$ \hat{V}_1 $}
\DashLine(-35,230)(0,200){2}
\Text(47,232)[]{$ \hat{V}_3 $}
\DashLine(0,200)(35,230){2}
\Text(-47,165)[]{$ V_{2\beta} $}
\Photon(-35,170)(0,200){1}{7}
\Text(47,165)[]{$ V_{4\delta} $}
\Photon(0,200)(35,170){1}{7}
\Text(90,200)[l]{$=~~\displaystyle ie^2\,C\,(-g_{\beta\delta}) $}
\end{picture}
\caption{\label{fig:3} 
Four-vector vertex (box on the top) and its corresponding $\epsilon$-vertices.}
\end{center}
\end{figure}

\begin{figure}[t]
\begin{center}
\fbox{
\begin{picture}(180,60)
\SetOffset(55,-175)
\Line(-35,200)(0,200)
\Text(-37,210)[]{$ S $}
\Photon(0,200)(25,215){1}{7}
\Text(47,225)[tr]{$ V_{1\alpha} $}
\Photon(25,190)(0,200){1}{7}
\Text(47,182)[br]{$  V_{2\beta} $}
\Text(55,200)[l]{$=~~\displaystyle ie\,C\,g_{\alpha\beta} $}
\end{picture}
}

\begin{picture}(180,70)
\SetOffset(55,-177)
\Line(-35,200)(0,200)
\Text(-37,210)[]{$ S $}
\DashLine(0,200)(25,215){2}
\Text(43,225)[tr]{$ \hat{V}_1 $}
\DashLine(25,190)(0,200){2}
\Text(43,182)[br]{$ \hat {V}_2 $}
\Text(55,200)[l]{$=~~\displaystyle ie\,C $}
\end{picture}
\caption{\label{fig:4} 
Two-vector-one-scalar vertex (box on the top) and its corresponding $\epsilon$-vertex.}
\end{center}
\end{figure}


\begin{figure}[t]
\begin{center}
\fbox{
\begin{picture}(210,60)
\SetOffset(55,-173)
\Photon(-35,200)(0,200){1}{7}
\Text(-35,210)[r]{$ V_{\alpha} $}
\Line(0,200)(25,215)
\Text(43,225)[tr]{$ S_1 $}
\Text(5,220)[l]{$ p_1 $}
\LongArrow(13,215)(5,210)
\Line(25,190)(0,200)
\Text(5,185)[l]{$ p_2 $}
\LongArrow(15,188)(5,192)
\Text(43,182)[br]{$ S_2 $}
\Text(55,200)[l]{$=~~\displaystyle ie\,C\,(p_1-p_2)_{\alpha} $}
\end{picture}
}

\begin{picture}(200,80)
\SetOffset(55,-175)
\DashLine(-35,200)(0,200){2}
\Text(-35,210)[r]{$ \hat{V} $}
\Line(0,200)(25,215)
\Text(43,225)[tr]{$ S_1 $}
\Text(0,222)[l]{$ \pm q $}
\LongArrow(13,215)(5,210)
\Line(25,190)(0,200)
\Text(43,182)[br]{$ S_2 $}
\Text(55,200)[l]{$=~~$}
\SetOffset(180,-175)
\DashLine(-35,200)(0,200){2}
\Text(-35,210)[r]{$ \hat{V} $}
\Line(0,200)(25,215)
\Text(43,225)[tr]{$ S_1 $}
\Line(25,190)(0,200)
\Text(0,182)[l]{$ \pm q $}
\LongArrow(5,192)(15,188)
\Text(43,182)[br]{$ S_2 $}
\Text(55,200)[l]{$=~~\displaystyle ie\,C\,(\pm i\mu) $}
\end{picture}
\caption{\label{fig:5} 
One-vector-two-scalar vertex (box on the top) and its corresponding $\epsilon$-vertex.
$q$ represents the flow of the loop momentum.
}\end{center}
\end{figure}


\begin{figure}[t]
\begin{center}
\fbox{
\begin{picture}(210,90)
\SetOffset(70,-155)
\Text(-47,235)[]{$ V_{1\alpha} $}
\Photon(-35,230)(0,200){1}{7}
\Text(47,232)[]{$ S_1  $}
\Line(0,200)(35,230)
\Text(-47,165)[]{$ V_{2\beta} $}
\Photon(-35,170)(0,200){1}{7}
\Text(47,165)[]{$ S_2 $}
\Line(0,200)(35,170)
\Text(70,200)[l]{$ =~~\displaystyle ie^2\,C\,g_{\alpha\beta}$}
\end{picture}
}

\begin{picture}(210,100)
\SetOffset(70,-155)
\Text(-47,235)[]{${\hat V}_1 $}
\DashLine(-35,230)(0,200){2}
\Text(47,232)[]{$ S_1 $}
\Line(0,200)(35,230)
\Text(-47,165)[]{$ \hat{V}_2 $}
\DashLine(-35,170)(0,200){2}
\Text(47,165)[]{$ S_2 $}
\Line(0,200)(35,170)
\Text(70,200)[l]{$=~~\displaystyle ie^2\,C $}
\end{picture}
\caption{\label{fig:6} 
Two-vector-two-scalar vertex (box on the top) and its corresponding $\epsilon$-vertex.
}\end{center}
\end{figure}

\section{\label{sec:4}Fermions and their interactions with vectors and scalars}
The study of the interactions involving fermions is complicated by the
presence of $\gamma_5$. It is convenient to start from the chiral fermions of the original theory and to split the fermion propagator 
in chirality flipping and chirality preserving parts, as illustrated 
in the top part of~\fig{fig:7}.  The $\epsilon$-propagator 
corresponding to the $\epsilon$-particle associated with a fermion is chirality flipping, being 
reminiscent of the presence of a $\rlap/ q$ in the numerator, as shown in 
the bottom part of~\fig{fig:7}.
Fermions can interact with vectors and scalars with the standard 
vector-fermion-fermion and scalar-fermion-fermion vertices shown
in the top parts of~\figss{fig:8}{fig:9}, while the
corresponding $\epsilon$-vertices are drawn in the bottom parts.  
Notice also that all vertices in~\fig{fig:8} are chirality flipping, 
due to the presence of $\gamma_\alpha$ in the original interaction,
while those in~\fig{fig:9} are chirality preserving, because of their 
scalar nature.
Particularly interesting is the vertex in ~\fig{fig:8} (c), which represents 
a $\hat V \hat F F$ interaction. Although no $\gamma$ matrix is present in that vertex, it should be considered as a chirality flipping one, because of its vectorial 
origin.

\begin{figure}[t]
\begin{center}
\begin{picture}(200,75)
\SetScale{0.6}
%
%
\sof(50,5)
\DashArrowLine(85,45)(107.5,67.5){3} \flin{107.5,67.5}{130,90} 
\flin{130,0}{107.5,22.5}
\DashArrowLine(107.5,22.5)(85,45){3}
\Photon(20,45)(85,45){2}{6}
\Photon(107.5,67.5)(107.5,22.5){2}{5}
\Text(60,45)[r]{\small $h$}
\Text(50,35)[r]{\small $-h$}
\Text(50,20)[r]{\small $h$}
\Text(60,12)[r]{\small $-h$}
\Text(67, 5)[r]{\small $h$}
\Text(8,29)[r]{\small $V_\alpha$}
\Text(81,54)[l]{$1$}
\Text(81,-2)[l]{$2$}
\end{picture}
\caption{\label{fig:example1}
$\epsilon$-diagram generating the fermionic structure 
$\bar v(1) \gamma_\beta \gamma_\alpha\gamma^\beta \omega^h u(2)$.}
\end{center} 
\end{figure}
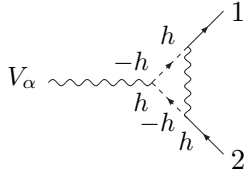

$\epsilon$-diagrams are built by using the rules of \figs{fig:7}{fig:9}
and reading, as usual, the fermionic line backward starting from the arrow. 
After the last vertex is encountered, a chirality projector  
$\omega^\pm = \frac{1}{2} (1\pm \gamma_5)$ should be inserted, according to the chirality of the fermion entering into it. For example, the diagram in~\fig{fig:example1} generates the fermionic structure
\bqa
\label{eq:ferm}
\bar v(1) \gamma_\beta \gamma_\alpha\gamma^\beta \omega^h u(2)\,.
\eqa
For fermionic loops, the starting point is arbitrary and should be kept fixed 
when summing over diagrams and families in order to preserve the right 
cancellations~\cite{Garzelli:2009is}.

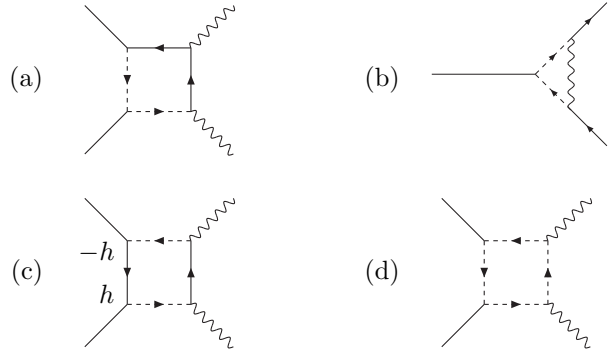
\begin{figure}[t]
\begin{center}
\begin{picture}(200,75)
\SetScale{0.6}
\sof(125,5)
\DashArrowLine(85,45)(107.5,67.5){3} \flin{107.5,67.5}{130,90} 
\flin{130,0}{107.5,22.5}
\DashArrowLine(107.5,22.5)(85,45){3}
\Line(20,45)(85,45)
\Photon(107.5,67.5)(107.5,22.5){2}{5}
\Text(0,25)[r]{\small (b)}

\SetScale{0.8}
\SetOffset(50,22)
\DashArrowLine(-35,-5)(-5,-5){2}
\ArrowLine(-5,25)(-35,25)
\DashArrowLine(-35,25)(-35,-5){2}
\ArrowLine(-5,-5)(-5,25)
\Line(-35,-5)(-55,-25)
\Photon(-5,-5)(15,-25){2}{5}
\Line(-35,25)(-55,45)
\Photon(-5,25)(15,45){2}{5}
\Text(-60,8)[r]{\small (a)}
\end{picture}

\begin{picture}(200,75)
\SetScale{0.8}
\SetOffset(50,25)
\DashArrowLine(-35,-5)(-5,-5){2}
\DashArrowLine(-5,25)(-35,25){2}
\ArrowLine(-35,25)(-35,-5) %
\ArrowLine(-5,-5)(-5,25)
\Line(-35,-5)(-55,-25)
\Photon(-5,-5)(15,-25){2}{5}
\Line(-35,25)(-55,45)
\Photon(-5,25)(15,45){2}{5}
\Text(-33,0)[r]{\small $h$}
\Text(-33,16)[r]{\small $-h$}
\Text(-60,8)[r]{\small (c)}
\SetOffset(185,25)
\DashArrowLine(-35,-5)(-5,-5){2}
\DashArrowLine(-5,25)(-35,25){2}
\DashArrowLine(-35,25)(-35,-5){2} %
\DashArrowLine(-5,-5)(-5,25){2}
\Line(-35,-5)(-55,-25)
\Photon(-5,-5)(15,-25){2}{5}
\Line(-35,25)(-55,45)
\Photon(-5,25)(15,45){2}{5}
\Text(-60,8)[r]{\small (d)}
\end{picture}

\caption{\label{fig:example2}
Fermionic $\epsilon$-diagrams. A minus sign should be assigned 
to the diagrams (a), (b) and (d), while no additional sign is required for (c).}
\end{center} 
\end{figure}

When scalars are present, the rules presented in \figs{fig:7}{fig:9}
define the diagrams up to a sign, which has to be included by hand.
This is due to the anticommutation properties of 
$\rlap/ {\tld{q}}$ 
with the four dimensional $\gamma$ matrices. The rule for fixing the sign is as follows. For each pair of $\epsilon$-fermion lines present in a given
diagram the result should be multiplied by $(-)^{(n_s+n_p)}$, where $n_s$ is the number of scalar vertices and $n_p$ the number of chirality preserving
propagators of the kind $i \frac{m_F}{p^2-m^2_F}$ between them.
For example, a minus sign should be assigned to the diagrams
in~\fig{fig:example2} (a), (b) and (d) whereas a plus sign should be given 
to that one in~\fig{fig:example2} (c).

\begin{figure}[t]
\begin{center}

\begin{picture}(200,75)
\SetScale{0.6}
%
%
\sof(40,5)
\DashArrowLine(85,45)(107.5,67.5){3} \flin{107.5,67.5}{130,90} 
\flin{130,0}{107.5,22.5}
\DashArrowLine(107.5,22.5)(85,45){3}
\Photon(20,45)(85,45){2}{6}
\DashLine(107.5,67.5)(107.5,22.5){3}
\end{picture}
\caption{\label{fig:example3}
Example of fermionic $\epsilon$-diagram that should be discarded because 
both ends of the $\epsilon$-gluon connect to a fermionic line.}
\end{center} 
\end{figure}
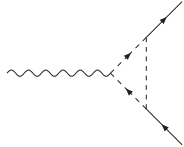

A last subtlety concerns again the vertex in~\fig{fig:8} (c). The
scalar $\epsilon$-dimensional degrees of freedom brought by 
the field $ \hat V$ can only originate from one of the vertices introduced 
in section~\ref{sec:3}. In fact, $\hat V$ fields are only needed for
non abelian theories because of the momentum dependent  
three-vector vertex of~\fig{fig:2}
\footnote{In pure QED, one just needs to introduce 
one $\hat F$ field for each fermion family.}. 
Therefore, diagrams where both ends of a $\hat V$ $\epsilon$-particle connect to a fermion line should be discarded, such as that one given
in~\fig{fig:example3}.

Finally, in the case of Majorana fermions, such as neutralinos and gluinos in SUSY theories, the relative sign of interfering Feynman diagrams can be determined as described in~\cite{Denner:1992vza,Denner:1992me}.  

\begin{figure}[t]
\begin{center}
\fbox{
\begin{picture}(410,90)
\SetOffset(65,55)
\ArrowLine(-50,10)(20,10)
\Text(-53,0)[l]{$F$}
\Text(12,0)[l]{$\overline{F} $}
\Text(-18,22)[l]{$p $}
\LongArrow(-28,15)(-3,15)
\Text(30,10)[l]{$\displaystyle = $}
\Text(55,10)[l]{$\displaystyle \frac{i}{\rlap/p - m_F} $}
\Text(105,10)[l]{$\displaystyle = $}
\SetOffset(255,55)
\ArrowLine(-60,10)(30,10)
\Text(-63,0)[l]{$ F $}
\Text(20,0)[l]{$\overline{F} $}
\Text(-60,16)[l]{$ -h $}
\Text(21,16)[l]{$ h $}
\Text(-16,22)[l]{$ p $}
\LongArrow(-25,15)(0,15)
\Text(40,10)[l]{$\displaystyle + $}
\SetOffset(366,55)
\ArrowLine(-50,10)(30,10)
\Text(-54,0)[l]{$ F $}
\Text(20,0)[l]{$\overline{F} $}
\Text(-50,16)[l]{$ h $}
\Text(21,16)[l]{$ h $}
\Text(-16,22)[l]{$ p $}
\LongArrow(-25,15)(0,15)
\end{picture}
}

\begin{picture}(210,90)
\SetOffset(-35,100)
\ArrowLine(-50,10)(20,10)
\Text(-52,16)[l]{$ -h $}
\Text(14,16)[l]{$ h $}
\Text(-53,0)[l]{$F$}
\Text(12,0)[l]{$\overline{F} $}
\Text(-18,22)[l]{$p $}
\LongArrow(-28,15)(-3,15)
\Text(30,10)[l]{$\displaystyle = $}
\Text(55,10)[l]{$\displaystyle i \frac{\rlap/p}{p^2-m^2_F} $,}
\SetOffset(165,100)
\ArrowLine(-50,10)(20,10)
\Text(-52,16)[l]{$~h$}
\Text(14,16)[l]{$ h $}
\Text(-53,0)[l]{$F$}
\Text(15,0)[l]{$\overline{F} $}
\Text(-18,22)[l]{$p $}
\LongArrow(-28,15)(-3,15)
\Text(30,9)[l]{$\displaystyle = $}
\Text(55,9)[l]{$\displaystyle i \frac{m_F}{p^2-m^2_F}\, $.}
\end{picture}

\begin{picture}(210,30)
\SetOffset(-35,100)
\SetOffset(0,80)
\DashArrowLine(-50,10)(30,10){2}
\Text(-54,0)[l]{$ \hat{F} $}
\Text(22,0)[l]{$ \hat{\overline{F}} $}
\Text(-16,22)[l]{$ p $}
\LongArrow(-25,15)(0,15)
\Text(45,10)[l]{$\displaystyle = $}
\Text(110,10)[r]{$\displaystyle  \frac{i}{p^2-M^2_F} $}
\Text(135,10)[l]{$\displaystyle = $}
\SetOffset(220,80)
\DashArrowLine(-50,10)(30,10){2}
\Text(-54,0)[l]{$ \hat{F} $}
\Text(22,0)[l]{$ \hat{\overline{F}} $}
\Text(-52,16)[l]{$ -h $}
\Text(21,16)[l]{$ h $}
\Text(-16,22)[l]{$ p $}
\LongArrow(-25,15)(0,15)
\end{picture}

\vspace{-2.5cm}

\caption{\label{fig:7} 
In the box on the top a fermion propagator is split in 
chirality flipping and chirality preserving parts.
$h= \pm$ denotes right-handed or left-handed
components and the dashed line on the bottom represents
the propagator of the $\epsilon$-particle associated with a fermion.
}
\end{center}
\end{figure}

\begin{figure}[t]
\begin{center}
\fbox{
\begin{picture}(200,60)
\SetOffset(53,-173)
\Photon(-35,200)(0,200){2}{7}
\Text(-35,212)[]{$V_{\alpha}$}
\ArrowLine(0,200)(25,215)
\Text(43,225)[tr]{$\overline{F}_1$}
\ArrowLine(25,190)(0,200)
\Text(8,212)[r]{$ -h $}
\Text(8,192)[r]{$ h $}
\Text(43,182)[br]{$F_2$}
\Text(55,202)[l]{$=~~\displaystyle ie{\gamma_{\alpha}}C^h{\omega_h} $}
\end{picture}
}

\begin{picture}(200,60)
\SetOffset(53,-173)
\Photon(-35,200)(0,200){2}{7}
\Text(-35,212)[]{$V_{\alpha}$}
\DashArrowLine(0,200)(25,215){2}
\Text(43,225)[tr]{$\hat{\overline{F}}_1$}
\DashArrowLine(25,190)(0,200){2}
\Text(8,212)[r]{$ -h $}
\Text(8,192)[r]{$ h $}
\Text(43,182)[br]{$\hat{F}_2$}
\Text(55,202)[l]{$=~~\displaystyle ie{\gamma_{\alpha}}C^h\mu $}
\Text(-100,202)[l]{(a)}
\end{picture}

\begin{picture}(200,60)
\SetOffset(53,-173)
\Photon(-35,200)(0,200){2}{7}
\Text(-35,212)[]{$V_{\alpha}$}
\DashArrowLine(0,200)(25,215){2}
\Text(43,225)[tr]{$\hat{\overline{F}}_1$}
\ArrowLine(25,190)(0,200)
\Text(8,212)[r]{$ -h $}
\Text(8,192)[r]{$ h $}
\Text(43,182)[br]{$ F_2 $}
\Text(55,202)[l]{$=$}
\Text(-100,202)[l]{(b)}
\SetOffset(170,-173)
\Photon(-35,200)(0,200){2}{7}
\Text(-35,210)[]{$V_{\alpha}$}
\ArrowLine(0,200)(25,215)
\Text(43,225)[tr]{$\overline{F}_1$}
\DashArrowLine(25,190)(0,200){2}
\Text(8,212)[r]{$ -h $}
\Text(8,192)[r]{$ h $}
\Text(43,182)[br]{$ \hat{F}_2 $}
\Text(55,202)[l]{$=~~\displaystyle ie{\gamma_{\alpha}}C^h \sqrt{\mu} $}
\end{picture}

\begin{picture}(200,60)
\SetOffset(53,-173)
\DashLine(-35,200)(0,200){2}
\Text(-35,212)[]{$\hat V $}
\ArrowLine(0,200)(25,215)
\Text(43,225)[tr]{$\overline{F}_1$}
\DashArrowLine(25,190)(0,200){2}
\Text(8,212)[r]{$ -h $}
\Text(8,192)[r]{$ h $}
\Text(43,182)[br]{$ \hat{F}_2 $}
\Text(12,181)[l]{$ \pm q $}
\LongArrow(22,186)(12,190)
\Text(55,202)[l]{$=$}
\Text(-100,202)[l]{(c)}
\SetOffset(170,-173)
\DashLine(-35,200)(0,200){2}
\Text(-35,210)[]{$\hat V$}
\DashArrowLine(0,200)(25,215){2}
\Text(43,225)[tr]{$\hat{\overline{F}}_1$}
\ArrowLine(25,190)(0,200)
\Text(8,212)[r]{$ -h $}
\Text(8,192)[r]{$ h $}
\Text(43,182)[br]{$ F_2 $}
\Text(5,225)[l]{$ \pm q $}
\LongArrow(13,214)(21,219)
\Text(55,202)[l]{$=~~\displaystyle ieC^h (\pm i\sqrt{\mu})$}
\end{picture}

\caption{\label{fig:8} 
Vector-fermion-fermion vertex (box on the top) and its corresponding $\epsilon$-vertices.
$h= \pm$ denotes right-handed or left-handed fermion components,
$\omega_h$ is a chirality projector and 
$q$ represents the flow of the loop momentum.}
\end{center}
\end{figure}

\begin{figure}[t]
\begin{center}
\fbox{
\begin{picture}(200,60)
\SetOffset(53,-173)
\Line(-35,200)(0,200)
\Text(-32,210)[]{$ S $}
\ArrowLine(0,200)(25,215)
\Text(43,225)[tr]{$\overline{F}_1$}
\ArrowLine(25,190)(0,200)
\Text(8,212)[r]{$ h $}
\Text(8,192)[r]{$ h $}
\Text(43,182)[br]{$F_2$}
\Text(55,202)[l]{$=~~\displaystyle ieC^h{\omega_h} $}
\end{picture}
}

\begin{picture}(200,60)
\SetOffset(53,-173)
\Line(-35,200)(0,200)
\Text(-32,210)[]{$ S $}
\DashArrowLine(0,200)(25,215){2}
\Text(43,225)[tr]{$\hat{\overline{F}}_1$}
\DashArrowLine(25,190)(0,200){2}
\Text(8,212)[r]{$ h $}
\Text(8,192)[r]{$ h $}
\Text(43,182)[br]{$\hat{F}_2$}
\Text(55,202)[l]{$=~~\displaystyle ieC^h\mu $}
\end{picture}

\begin{picture}(200,60)
\SetOffset(53,-173)
\Line(-35,200)(0,200)
\Text(-32,210)[]{$ S $}
\DashArrowLine(0,200)(25,215){2}
\Text(43,225)[tr]{$\hat{\overline{F}}_1$}
\ArrowLine(25,190)(0,200)
\Text(8,212)[r]{$ h $}
\Text(8,192)[r]{$ h $}
\Text(43,182)[br]{$ F_2 $}
\Text(55,202)[l]{$=$}
\SetOffset(170,-173)
\Line(-35,200)(0,200)
\Text(-32,210)[]{$ S $}
\ArrowLine(0,200)(25,215)
\Text(43,225)[tr]{$\overline{F}_1$}
\DashArrowLine(25,190)(0,200){2}
\Text(8,212)[r]{$ h $}
\Text(8,192)[r]{$ h $}
\Text(43,182)[br]{$ \hat{F}_2 $}
\Text(55,202)[l]{$=~~\displaystyle ieC^h \sqrt{\mu} $}
\end{picture}

\end{center}

\caption{\label{fig:9} 
Scalar-fermion-fermion vertex (box on the top) and its corresponding $\epsilon$-vertices.
$h= \pm$ denotes right-handed or left-handed fermion components and
$\omega_h$ is a chirality projector.}

\end{figure}

\section{\label{sec:5}QCD}
In the case of QCD no $\gamma_5$ is present and splitting fermions 
into right-handed and left-handed components is no longer necessary.
No additional sign needs to be inserted, since no scalar particles 
are involved in the $\epsilon$-fermionic vertices, although diagrams
in which a scalar $\epsilon$-gluon connects 2 fermionic lines 
should be discarded, as explained in the previous section. 
I list the relevant QCD $\epsilon$-propagators and 
$\epsilon$-vertices in~\fig{fig:10}. As can be seen, they have exactly the 
same general Lorentz structure described in sections~\ref{sec:3} and~\ref{sec:4}, while terms of the original color structures split among the various contributions in a well defined way.

\begin{figure}[t]
\begin{center}
\fbox{
\begin{picture}(410,250)
\SetScale{0.5}
%
%
\sof(-5,180)
\gluon{35,50}{135,50}{9}
\LongArrow(75,62)(95,62)
\Text(40,34)[bl]{$p$}
\Text(17,15)[]{$\alpha$,$a$}
\Text(68,15)[]{$\beta$,$b$}
\Text(75,25)[l]{$\displaystyle =- i\, \frac{g_{\alpha\beta}}{p^2}\, \delta_{ab}$ ,}
%
%
\sof(138,180)
\LongArrow(75,62)(98,62)
\Text(40,34)[bl]{$p$}
\flin{35,50}{135,50}
\Text(17,17)[]{$l$}
\Text(68,17)[]{$k$}
\Text(75,25)[l]{$\displaystyle = \frac{i\,\delta_{kl}}{\rlap/p - m_q}$ ,}
%
%
\sof(276,180)
\gluon{130,5}{85,50}{5}\flin{85,50}{130,95} \flin{20,50}{85,50}
\Text(70,55)[l]{$k$}
\Text(7,15)[l]{$l$}
\Text(75,-6)[]{$\alpha$,$a$}
\Text(75,25)[l]{$\displaystyle = -i g t_{kl}^a \gamma^{\alpha}\,$,}
%
%
\sof(5,100)
\gluon{130,5}{85,50}{5}\gluon{85,50}{130,95}{5}
\gluon{20,50}{85,50}{5}
\LongArrow(35,62)(66,62)
\LongArrow(105,10)(85,30)
\LongArrow(110,95)(90,75)
\Text(46,49)[]{$p_2$}
\Text(25,40)[]{$p_1$}
\Text(45,3)[]{$p_3$}
\Text(75,55)[]{$\beta$,$b$}
\Text(7,15)[]{$\alpha$,$a$}
\Text(75,-6)[]{$\gamma$,$c$}
\Text(90,25)[l]{$ \displaystyle  =  
g \,f^{abc}\,
\left[g_{\alpha\beta} (p_2-p_1)_\gamma
+g_{\beta\gamma} (p_3-p_2)_\alpha
+g_{\gamma\alpha}(p_1-p_3)_\beta
\right]\,$
,}
%
%
\sof(5,20)
\gluon{130,5}{85,50}{5}\gluon{85,50}{130,95}{5}
\gluon{40,5}{85,50}{5}\gluon{85,50}{40,95}{5}
\Text(68,0)[l]{$\gamma$,$c$}
\Text(17,0)[r]{ $\delta$,$d$}
\Text(68,50)[l]{ $\beta$,$b$}
\Text(17,50)[r]{ $\alpha$,$a$}
\Text(90,25)[l]{$\displaystyle = - ig^2\,[~
 f^{ebc}f^{eda}(g_{\beta \delta} g_{\alpha \gamma} 
             - g_{\alpha \beta} g_{\gamma\delta})
+f^{ebd}f^{eac}(g_{\alpha \beta} g_{\gamma \delta} 
             - g_{\beta\gamma} g_{\alpha \delta})$}
\Text(123,0)[l]{$ \displaystyle 
+f^{eba}f^{ecd}(g_{\beta \gamma} g_{\alpha \delta} 
              - g_{\beta \delta} g_{\alpha \gamma})~]\,.$}
\end{picture}
}

\begin{picture}(300,310)

\SetOffset(-10,60)
\DashLine(-35,200)(-5,200){2}
\LongArrow(-27,205)(-14,205)
\Text(-20,211)[]{$p$}
\Text(-35,191)[br]{$a$}
\Text(-5,191)[bl]{$b$}
\Text(5,201)[l]{$=~\displaystyle -i \frac {\delta_{ab}}{p^2}\,$,}

\SetOffset(105,60)
\DashArrowLine(-35,200)(-5,200){2}
\LongArrow(-27,205)(-14,205)
\Text(-20,211)[]{$p$}
\Text(-35,191)[br]{$l$}
\Text(-5,191)[bl]{$k$}
\Text(5,201)[l]{$=~\displaystyle i \frac {\delta_{kl}}{p^2-m_q^2}\,$,}

\SetOffset(230,60)
\Gluon(-35,200)(0,200){3}{5}
\Text(-25,190)[r]{$\alpha$,$a$}
\DashArrowLine(0,200)(25,215){2}
\Text(35,222)[tr]{$k$}
\DashArrowLine(25,190)(0,200){2}
\Text(33,185)[br]{$l$}
\Text(35,201)[l]{$=~\displaystyle  - ig\,{\mu}\, t_{kl}^a  
\,{\gamma_{\alpha}}\,$,}

\SetOffset(-10,0)
\DashLine(-35,200)(0,200){2}
\Text(-35,193)[]{$a$}
\ArrowLine(0,200)(25,215)
\Text(35,222)[tr]{$k$}
\DashArrowLine(25,190)(0,200){2}
\Text(34,185)[br]{$l$}
\LongArrow(18,188.2)(7,193)
\Text(15,183)[r]{$\pm q$}
\Text(40,201)[l]{$\displaystyle  = $}

\SetOffset(110,0)
\DashLine(-35,200)(0,200){2}
\Text(-35,193)[]{$a$}
\DashArrowLine(0,200)(25,215){2}
\Text(35,222)[tr]{$k$}
\ArrowLine(25,190)(0,200)
\Text(33,185)[br]{$l$}
\LongArrow(8,210)(18,216)
\Text(13,220)[r]{$\pm q$}
\Text(40,201)[l]{$\displaystyle  = - ig\,(\pm i \sqrt{\mu}) \,t_{kl}^a\,$,}

\SetOffset(-10,-60)
\Gluon(-35,200)(0,200){3}{5}
\Text(-25,190)[r]{$\alpha$,$a$}
\ArrowLine(0,200)(25,215)
\Text(35,222)[tr]{$k$}
\DashArrowLine(25,190)(0,200){2}
\Text(35,185)[br]{$l$}
\Text(40,201)[l]{$\displaystyle =$}

\SetOffset(110,-60)
\Gluon(-35,200)(0,200){3}{5}
\Text(-25,190)[r]{$\alpha$,$a$}
\DashArrowLine(0,200)(25,215){2}
\Text(35,222)[tr]{$k$}
\ArrowLine(25,190)(0,200)
\Text(33,185)[br]{$l$}
\Text(40,201)[l]{$\displaystyle  = - ig\,\sqrt{\mu}\, t_{kl}^a  \,{\gamma_{\alpha}}\,$,}

\SetOffset(-10,-120)
\DashLine(-35,200)(0,200){2}
\Text(-35,195)[]{$a$}
\Gluon(25,215)(0,200){-3}{4}
\Text(35,222)[]{$\beta$,$b$}
\Gluon(0,200)(25,190){3}{4}
\Text(33,185)[]{$\gamma$,$c$}
\LongArrow(18,185.2)(7,190)
\Text(15,181)[r]{$\pm q$}
\Text(40,201)[l]{$\displaystyle =$}

\SetOffset(110,-120)
\DashLine(-35,200)(0,200){2}
\Text(-35,195)[]{$a$}
\Gluon(25,215)(0,200){-3}{4}
\Text(35,222)[]{$\beta$,$b$}
\Gluon(0,200)(25,190){3}{4}
\Text(33,185)[]{$\gamma$,$c$}
\LongArrow(8,213)(18,219)
\Text(13,222)[r]{$\pm q$}
\Text(40,201)[l]{$\displaystyle  = g\,(\pm i \mu)\, f^{abc} \,{g_{\beta\gamma}}\,$,}

\SetOffset(-10,-180)
\Gluon(-35,200)(0,200){3}{5}
\Text(-25,190)[r]{$\alpha$,$a$}
\DashLine(0,200)(25,215){2}
\Text(35,222)[]{$b$}
\LongArrow(18,215)(7,208)
\Text(15,220)[r]{$p_b$}
\DashLine(25,190)(0,200){2}
\Text(33,185)[]{$c$}
\LongArrow(18,189)(7,193)
\Text(15,184)[r]{$p_c$}
\Text(40,201)[l]{$\displaystyle  = gf^{abc}(p_c - p_b)_{\alpha}\,$,}

\SetOffset(180,-162)
\Gluon(-35,200)(-10,180){3}{4}
\Text(-38,205)[r]{$\alpha$,$a$}
\Gluon(-10,180)(15,200){3}{4}
\Text(33,205)[r]{$\beta$,$b$}
\DashLine(-35,160)(-10,180){2}
\Text(-39,155)[r]{$d$}
\DashLine(-10,180)(15,160){2}
\Text(23,153)[r]{$c$}
\Text(25,180)[l]{$=~\displaystyle -ig^2g_{\alpha\beta}(f^{ead}f^{ebc}+f^{eac}f^{ebd})\,$.}
\end{picture}
\end{center}
\caption{
\label{fig:10} 
QCD Feynman rules (box on the top) and corresponding
$\epsilon$-propagators and $\epsilon$-vertices.
$q$ represents the flow of the loop momentum.
}
\end{figure}

\section{\label{sec:6}Conclusions}
I presented the set of special Feynman rules allowing the reconstruction
of the $\epsilon$-dimensional part of 1-loop amplitudes in theories with vectors, scalars and fermions.
The rules are quite simple, when assuming a renormalizable gauge, and easily derivable from the vertices of the original theory. 
They can be used to extract the $\mu^2$ dependence from the 
integrand of any contributing 1-loop Feynman diagram, namely  the
$\epsilon$-dimensional part generated by self contractions of the loop momentum.

The complete electroweak model can be studies by simply fixing the constants 
appearing in the vertices of \figs{fig:1}{fig:6}
and~\figs{fig:7}{fig:9} to their standard model values, while
the interactions relevant for QCD are explicitly listed in~\fig{fig:10}.

A four dimensional helicity scheme is used in this work, but simple 
translation rules to the 't Hooft Veltman scheme are collected in an appendix.

SUSY and BSM theories sharing the same Lorentz structures studied in this paper
can be treated in the same way.

The special vertices presented here may also be considered as a 
practical tool to determine the counter-terms needed to restore gauge invariance in calculations where the numerator function of the 1-loop Feynman diagrams is computed in four dimensions. This possibility is particularly appealing in conjunction with schemes such as dimensional reduction, where the use of particular classes of four-dimensional identities involving $\gamma_5$ is forbidden.

It would also be interesting to generalize this approach to the Unitary 
gauge and beyond 1-loop. I leave these two issues to future investigations.

\section*{Acknowledgments}
I would like to thank Fabio Maltoni for useful discussions.
This research is supported by the MEC project FPA2008-02984.

\appendix
\section{\label{sec:app}From the FDH scheme to the HV scheme}
In this appendix, I give the translation rules from the FDH scheme (or dimensional reduction) reproduced  by the Feynman rules given 
in sections~\ref{sec:3} to~\ref{sec:5} and
the HV scheme of \eqn{eqr2}.
 For QCD, once a 1-loop amplitude $A^{(1)}$ has been computed in FDH, the 
corresponding HV result can be obtained with the help of the formula~\cite{signerth}
\bqa
A^{(1)}_{\rm HV} =  A^{(1)}_{\rm FDH} + A^{(0)} \frac{g^2}{16{\pi^2}} \Bigl[\, \frac{N_c}{6}(n_q+n_Q-2)- \frac {n_q}{2} \, \frac{N^2_c-1}{2N_c}   \Bigr ]\,,
\eqa
where $A^{(0)}$ is the tree level result, $N_c$ 
the number of colors, $n_q$ the number of massless quarks and
$n_Q$ the number of massive quarks. 
For the electroweak model, once the renormalized 1-loop amplitude has been 
determined in FDH, since the terms proportional to 
$\epsilon$ in~\eqn{eqr2} are separately gauge invariant, their total
contribution can be completely reabsorbed by shifts of 
the renormalization constants.
The HV result can then be obtained through the replacements
\footnote{I use the 
same notations and conventions of~\cite{Denner:1991kt} and assume a unit CKM matrix.}
\bqa
\label{ewscheme}
\begin{tabular}{lcll} 
 $\delta t$  & $\to$  & $\delta t$  & $\displaystyle - \frac{e}{8 \pi^2s}M^3_W\left(1+\frac{1}{2 c^4}\right) $  \\ \vspace{-0.32cm} \\
 $\delta M^2_H $  & $\to$  & $\delta M^2_H $  & $\displaystyle+3\frac{e^2}{16 \pi^2 s^2} M^2_W\left(1+\frac{1}{2 c^4}\right)$  \\\vspace{-0.32cm}\\
 $\delta Z_H $  & $\to$  & $\delta Z_H $  & $$  \\\vspace{-0.32cm}\\
 $\delta M^2_W $  & $\to$  & $\delta M^2_W $  & $\displaystyle + \frac{e^2}{24 \pi^2 s^2} M^2_W$  \\\vspace{-0.32cm}\\
 $\delta Z_W $  & $\to$  & $\delta Z_W $  & $\displaystyle - \frac{e^2}{24 \pi^2 s^2}$  \\\vspace{-0.32cm}\\
 $\delta M^2_Z $  & $\to$  & $\delta M^2_Z $  & $\displaystyle + \frac{e^2 c^2}{24 \pi^2 s^2} M^2_Z$  \\\vspace{-0.32cm}\\
 $\delta Z_{ZZ}$  & $\to$  & $\delta Z_{ZZ}$  & $\displaystyle   - \frac{e^2 c^2}{24 \pi^2 s^2} $  \\\vspace{-0.32cm}\\
 $\delta Z_{AZ}$  & $\to$  & $\delta Z_{AZ}$  & $\displaystyle + \frac{e^2 c}{12 \pi^2 s} $  \\\vspace{-0.32cm}\\
 $\delta Z_{ZA} $  & $\to$  & $\delta Z_{ZA} $  & $\displaystyle$  \\\vspace{-0.32cm}\\
 $\delta Z_{AA}$  & $\to$  & $\delta Z_{AA}$  & $\displaystyle  - \frac{e^2}{24 \pi^2} $  \\\vspace{-0.32cm}\\
 $\delta m_{f,i}$  & $\to$  & $\delta m_{f,i}$  & $\displaystyle  -\frac{m_{f,i}}{2} \frac{e^2}{16 \pi^2}
\left(
\frac{1}{4 s^2c^2}-6 \frac{Q_f I^3_{W,f}}{c^2}
+ 6 \frac{Q_f^2}{c^2}+\frac{1}{2 s^2}
\right)$  \\\vspace{-0.32cm}\\
 $\delta Z^{f,L}_{ii}$  & $\to$  & $\delta Z^{f,L}_{ii}$  & $\displaystyle 
+\frac{e^2}{16 \pi^2}
\left(
\frac{1}{4 s^2c^2}-2 \frac{Q_f I^3_{W,f}}{c^2}
+ \frac{Q_f^2}{c^2}+\frac{1}{2 s^2}
\right) $  \\\vspace{-0.32cm}\\
 $\delta Z^{f,R}_{ii}$  & $\to$  & $\delta Z^{f,R}_{ii}$  & $\displaystyle + \frac{e^2}{16 \pi^2} \frac{Q^2_f}{c^2}  \,.$  
\end{tabular}
\eqa
From \eqn{ewscheme}, the necessary shifts in the charge renormalization constant and in  the sine and cosine of the weak mixing angle (relevant when
using the on-shell scheme) can be determined from the equations
\bqa
\delta Z_e &=& -\frac{1}{2} \left( \delta Z_{AA}+ \frac{s}{c} \delta Z_{ZA} 
\right) \nl
\frac{\delta c}{c} &=& \frac{1}{2} \left(\frac{\delta M^2_W}{M^2_W}
-\frac{\delta M^2_Z}{M^2_Z} \right) \nl
\frac{\delta s}{s} &=& -\frac{c^2}{s^2}\frac{\delta c}{c}\,.
\eqa
The rules in \eqn{ewscheme} are easily derived from the explicit  
knowledge of the part proportional to $\lambda_{\rm HV}$ in the
2-point functions listed in~\cite{Garzelli:2009is}.


\bibliographystyle{JHEP}	
\bibliography{feyn}		

\end{document}